\documentclass[fleqn,12pt,twoside]{article}

\usepackage{espcrc1}

\usepackage{graphicx}

\usepackage[figuresright]{rotating}

\title{Nuclear Structure Calculations with Coupled Cluster Methods 
from Quantum Chemistry}
\author{
D.J.~Dean\address[ORNL]{Physics Division, Oak Ridge National Laboratory,
P.O. Box 2008, Oak Ridge, TN 37831, USA}\address[OSLO1]{Center of Mathematics for Applications, 
University of Oslo, N-0316 Oslo, Norway},
J.R.~Gour\address[MSUCEM]{Department of Chemistry, Michigan State University, East Lansing, MI 48824, USA},
G.~Hagen\addressmark[OSLO1],
M.~Hjorth-Jensen\addressmark[OSLO1]\address[OSLO2]{Department of Physics, 
University of Oslo, N-0316 Oslo, Norway}\address[CERN]{PH Division, CERN, 
CH-1211 Geneve 23, Switzerland}\address[MSUPH]{Department of Physics and Astronomy, 
Michigan State University, East Lansing, MI 48824, USA}, 
K.~Kowalski\addressmark[MSUCEM], 
T.~Papenbrock\addressmark[ORNL]\address[UTK]{Department of Physics and Astronomy, University of 
Tennessee, Knoxville, TN 37996, USA}, 
P.~Piecuch\addressmark[MSUCEM]\addressmark[MSUPH], and M.~W{\l}och\addressmark[MSUCEM]} 
       
\begin{document}

\maketitle

\begin{abstract}
We present several coupled-cluster calculations of ground and excited states 
of $^4$He and $^{16}$O employing methods from quantum chemistry. 
A comparison
of coupled cluster results with the results of exact
diagonalization of the hamiltonian in the same model space and 
other truncated shell-model calculations shows that
the quantum chemistry inspired coupled cluster approximations provide
an excellent description of ground and excited states of nuclei, 
with much less computational
effort than traditional large-scale shell-model approaches. Unless truncations are made, 
for nuclei like  $^{16}$O,
full-fledged shell-model calculations with four or more major shells are not possible.
However, these and even larger systems can be studied with the coupled cluster methods
due to the polynomial rather than factorial scaling inherent in standard shell-model
studies. 
This makes the coupled cluster approaches, developed in quantum chemistry, viable methods 
for describing weakly bound systems of interest
for future nuclear facilities.
\end{abstract}

\section{INTRODUCTION}

Physical properties, such as masses and life-times,
of very short-lived, and hence very rare, nuclei are important
ingredients that determine element production mechanisms in
the universe. Given that present and future nuclear structure research facilities
will open significant
territory into regions of medium-mass and heavier nuclei,
it becomes important to investigate theoretical methods that will allow
for a description of medium-mass systems that are involved in such
element production. Such systems pose significant
challenges to existing nuclear structure models, especially since many of
these nuclei will be unstable and short-lived. How to deal with weakly
bound systems and coupling to resonant states is an unsettled problem in
nuclear spectroscopy. Furthermore, existing shell-model methods  are limited to
very few major shells and/or number of active particles. Similar constraints apply to 
ab initio Monte Carlo approaches. It is thus critical to develop new 
computational techniques which not only can handle several major 
shells but also perform ab initio
calculations starting with the free nucleon-nucleon interaction in many-body
systems larger than for example $^{40}$Ca.

In this work we focus on coupled cluster methods in our 
discussion of systems involving many 
single-particle degrees of freedom. 
The ab initio coupled-cluster theory is a particularly promising
candidate for such endeavors due to its enormous success in quantum
chemistry. 
Based on the experience from quantum chemistry, where coupled cluster methods 
can be applied to large molecular systems with more than hundred  
correlated electrons, we anticipate that quantum chemistry inspired coupled cluster
approaches will enable accurate studies of  ground and 
excited states of nuclei
with dimensionalities beyond the capability of present shell-model
approaches, with a much smaller numerical effort.
Even though the shell-model combined with appropriate effective interactions
offers in general a very good description of several stable and even weakly 
bound nuclei, the increasing single-particle level density of weakly bound systems
makes it imperative to 
identify and investigate methods that will extend to unstable systems,
systems whose dimensionality is beyond reach for present shell-model studies, 
typically limited today to  systems with at most $\sim 10^9$ basis states.

In this contribution we present results of coupled cluster calculations 
for ground and excited states of $^{4}$He and
$^{16}$O. Where possible, we compare these calculations with exact diagonalization from 
shell-model studies
within the same model spaces and with the same interaction. This serves to underline the reliability of the 
coupled cluster method in the nuclear many-body problem. 
The coupled cluster calculations are rather 
inexpensive compared with the shell-model approach, a feature which is very useful if 
one wants to include additional degrees of freedom such as more single particle levels. We end this contribution 
with a discussion of future projects.

\section{COUPLED CLUSTER APPROACH TO NUCLEI}

Coupled cluster theory originated in nuclear physics
\cite{coester58} around 1960.  Early studies in the
seventies \cite{klz78} probed ground-state properties in limited
spaces with free nucleon-nucleon interactions available at the
time. The subject was revisited
only recently by Bishop {\it et al.}
\cite{ticcm}, for further theoretical development, and by Mihaila and
Heisenberg \cite{hm99}, for coupled cluster calculations
using realistic
two- and three-nucleon
bare interactions
and expansions in the
inverse particle-hole energy spacings.
However, much of
the impressive development in
coupled cluster theory made in quantum chemistry in
the last 20-25 years
\cite{comp_chem_rev00,Bartlett95,paldus,cizek,Piecuch02a,Piecuch02b},
after the introduction of coupled-cluster theory and diagrammatic
methods to chemistry, by {\v C}{\i}{\v z}ek and Paldus \cite{paldus,cizek}, 
still awaits applications to the nuclear many-body problem.

Many solid theoretical reasons exist that motivate a pursuit of
coupled-cluster methods. First of all, the method is fully
microscopic and is capable of systematic and hierarchical improvements.
Indeed, when one expands the cluster operator in coupled-cluster theory
to all $A$ particles in the system, one exactly produces the fully-correlated
many-body wave function of the system. The only input that the method
requires is the nucleon-nucleon interaction. 
The method may also be extended
to higher-order interactions such as the three-nucleon interaction.
Second, in its standard formulation, 
the method is size extensive meaning that only linked
diagrams appear in the computation of the  
energy (the expectation value of the Hamiltonian) and amplitude equations.
It is well-known in for example quantum chemistry that all shell model calculations
that use particle-hole truncation schemes
actually suffer from the inclusion of disconnected diagrams
in computations of the energy.
Third, coupled-cluster theory is also size
consistent which means that the energy of two non-interacting fragments
computed separately is the same as that computed for both fragments
simultaneously. In chemistry, where the study of reactive and non-reactive collisions
of molecules are very important, this is a crucial property not available
in the truncated shell model (named limited configuration interaction in
chemistry).
Fourth, while the theory
is not variational, it does not have a bound,
the energy behaves as a variational quantity in most instances.
Finally, from a
computational point of view, the practical implementation of coupled
cluster theory is amenable to parallel computing. 

The basic idea of coupled-cluster theory is that the correlated many-body
wave function $\mid \Psi\rangle$ 
may be obtained by application of a cluster operator, 
$T$, such that
$|\Psi \rangle =\exp\left(T\right)|\Phi\rangle$
where $\Phi$ is a reference Slater determinant chosen as a convenient starting
point.  For example, we use the filled $0s$ state as the reference 
determinant for $^4$He.

The cluster operator $T$ is given by $T=T_1 + T_2 + \cdots T_A$ 
and represents various
$n$-particle-$n$-hole ($n$p-$n$h) excitation amplitudes such as
\begin{equation}
T_1 = \sum_{a>\varepsilon_f, i\le\varepsilon_f}t^i_a a^\dagger_a a_i \hspace{1cm}
T_2 = \frac{1}{4}\sum_{i,j\le\varepsilon_f; ab>\varepsilon_f}t^{ij}_{ab}
a^\dagger_a a^\dagger_b a_j a_i,
\end{equation}
and higher-order terms for $T_3$ to $T_A$.  
The basic approximation is obtained by truncating the many-body
expansion of $T$ at the $2p-2h$ cluster component $T_{2}$.
This is
commonly referred to in the literature as coupled-cluster singles and
doubles (CCSD).
We compute the ground-state energy from
\begin{equation}
E_{0}=\langle\Phi|\bar{H}^{{\rm (CCSD)}}|\Phi\rangle,
\end{equation} where 
$\bar{H}^{{\rm (CCSD)}}=\exp\left(-T\right) H \exp\left(T\right)$ is the coupled cluster similarity
transformed hamiltonian. In CCSD we set $T=T_1+T_2$.
To derive the CCSD or other coupled cluster approaches, we use the diagrammatic approach.
In order to obtain the computationally efficient algorithms,
which lead to the lowest operation count and memory requirements,
it is better to use the idea of recursively generated intermediates and
diagram factorization \cite{Bartlett95}.
The resulting equations can be cast into a computationally
efficient form, where diagrams representing intermediates multiply
diagrams representing cluster operators. The resulting equations
can be solved using efficient iterative algorithms, 
see for example Refs.~\cite{Bartlett95,ref1}.

\subsection{Effective Two-body Hamiltonian}

In shell-model studies of various nuclear mass regions, the common approach 
in deriving an effective interaction for the shell-model has been to start
with various perturbative many-body approaches, for recent reviews 
see Refs.~\cite{mhj95,dehko04}.

The starting point for the derivation of such an effective interaction is normally a
bare nucleon-nucleon interaction fitted to reproduce low-energy scattering data.
However, since this interaction has a strongly repulsive core at short internucleon 
distances, one needs to renormalize the short-range part in order to render
it suitable for an eventual perturbative treatment. To do that, one sums normally
the class of two-body particle-particle ladder diagrams to infinite order. 
This yields 
a new and renormalized interaction, the so-called reaction matrix $G$ or just 
the $G$-matrix. It is however energy dependent, as is the scattering matrix $T$.
It differs from the free scattering matrix $T$ by the introduction of a Pauli operator
accounting for a specific nuclear medium. 
The $G$-matrix is in turn used in a perturbative many-body scheme including higher-order
corrections, such as core-polarization terms. 
Such an effective interaction has 
been a very succesful starting point for shell-model studies. 
To derive effective interactions within
the framework of many-body perturbation theory
is however hard to expand upon in a systematic manner by including for example
three-body diagrams.
In addition, there are
no clear signs of convergence, even  in terms of a weak interaction 
such as the $G$-matrix. Even in atomic
 and molecular physics, many-body perturbative
methods are not much favoured any longer, see for example Ref.~\cite{helgaker} for a
critical discussion.
The lessons from atomic and molecular many-body systems clearly point to
the need of non-perturbative resummation techniques of large
classes of diagrams.

This is one of the main reasons for why we 
have chosen to focus on the coupled cluster method. 
However, coupled cluster calculations of nuclei, 
see Refs.~\cite{klz78,ticcm,hm99} have typically started
with the bare nucleon-nucleon interaction. As mentioned above, 
to renormalize this interaction one needs a 
very large set of single-particle 
states. The latter makes our use of quantum chemistry algorithms of 
little practical use if we were to start
with the bare interaction.

To circumvent this problem, we define an effective two-body 
hamiltonian taylored to a specific
model space. The single-particle states 
defining the model space, are in turn used as the basis for our coupled cluster
calculations. Here we employ a $G$-matrix defined with a so-called no-core 
Pauli operator, with
the harmonic oscillator defining our single-particle basis. 
Our model space is then a function of
various harmonic oscillator shells. The two-body states defining the 
$G$-matrix model space are  shown in Fig.~ \ref{fig:paulioperator}, 
with $n_3$ representing a large number, at least eight to ten major
oscillator shells.   
The single-particle states labeled by $n_3$ 
represent then the last orbit of the model space $P$, 
This so-called no-core model space is used in our definitions of 
model spaces for the resummations of many-body terms in 
coupled cluster theory. 
In Fig.~\ref{fig:paulioperator} the two-body state
$\left| (pq)JT_Z\right \rangle$ 
does not belong to the model space and is included
in the computation of
the $G$-matrix.
Similarly,
$\left| (p\gamma)JT_Z\right \rangle$
and
$\left| (\delta q)JT_Z\right \rangle$
also enter the definition of $Q$ whereas
$\left| (\delta\gamma)JT_Z\right \rangle$
is not included in the computation of $G$.
This means that correlations not defined in the $G$-matrix need
to be computed by other non-perturbative
resummations or many-body schemes.
This is where the coupled-cluster scheme enters.

With the $G$-matrix model space $P$ of Fig.~\ref{fig:paulioperator} we can now define
an appropriate space for 
coupled-cluster calculations where correlations
not included in the $G$-matrix are to be generated. This model space is defined
in Fig.~\ref{fig:finalp}, where the label $n_{p}$ represents the same
single-particle orbit as $n_{3}$ in  Fig.~\ref{fig:paulioperator}.

The $G$-matrix computed according to Fig.~\ref{fig:paulioperator}
does not reflect a specific nucleus and
thereby single-particle orbits which define the uncorrelated
Slater determinant.  For a nucleus like
$^{4}$He the $0s_{1/2}$ orbit is fully occupied and defines thereby single-hole states.
These are labeled by $n_{\alpha}$ in Fig.~\ref{fig:finalp}.
For $^{16}$O the corresponding hole states are represented by the orbits
$0s_{1/2}$,  $0p_{3/2}$ and  $0p_{1/2}$. With this caveat we can then generate
correlations not included in the $G$-matrix and perform resummations of larger 
classes of diagrams.

\begin{figure}[htb]
\begin{minipage}[t]{80mm}
%\framebox[79mm]{\rule[-26mm]{0mm}{52mm}}
\setlength{\unitlength}{0.6cm}
\begin{picture}(9,10)
\thicklines
   \put(1,0.5){\makebox(0,0)[bl]{
              \put(0,1){\vector(1,0){8}}
              \put(0,1){\vector(0,1){8}}
              \put(7,0.5){\makebox(0,0){${q}$}}
              \put(7,0.7){\line(0,1){0.3}}
              \put(2,0.5){\makebox(0,0){${\gamma}$}}
              \put(2,0.7){\line(0,1){0.3}}
              \put(-0.6,6){\makebox(0,0){$n_3$}}
              \put(5,0.5){\makebox(0,0){$n_3$}}
              \put(2,3){\makebox(0,0){$\hspace{1cm}Q=0$}}
              \put(2,7){\makebox(0,0){$\hspace{1cm}Q=1$}}
              \put(-0.6,7.8){\makebox(0,0){${p}$}}
              \put(-0.3,8){\line(1,0){0.3}}
              \put(-0.6,1.8){\makebox(0,0){${\delta}$}}
              \put(-0.3,2){\line(1,0){0.3}}
              \put(0,6){\line(1,0){5}}
              \put(5,1){\line(0,1){5}}
         }}
\end{picture}
\caption{Definition of the exclusion operator $Q$ used to compute the $G$-matrix for large
spaces.\label{fig:paulioperator}}
\end{minipage}
\hspace{\fill}
\begin{minipage}[t]{75mm}
%\framebox[74mm]{\rule[-26mm]{0mm}{52mm}}
\setlength{\unitlength}{0.6cm}
\begin{picture}(9,10)
\thicklines
   \put(1,0.5){\makebox(0,0)[bl]{
              \put(0,1){\vector(1,0){8}}
              \put(0,1){\vector(0,1){8}}
              \put(7,0.5){\makebox(0,0){b}}
              \put(-0.6,6){\makebox(0,0){$n_p$}}
              \put(5,0.5){\makebox(0,0){$n_p$}}
              \put(-0.6,3){\makebox(0,0){$n_{\alpha}$}}
              \put(2,0.5){\makebox(0,0){$n_{\alpha}$}}
              \put(-0.6,8){\makebox(0,0){a}}
              \put(0,6){\line(1,0){5}}
              \put(5,1){\line(0,1){5}}
              \put(0,3){\line(1,0){2}}
              \put(2,1){\line(0,1){2}}
         }}
\end{picture}
\caption{Definition of particle and hole states
for coupled-cluster calculations
in large spaces.
The orbit represented by 
 $n_{\alpha}$ stands for the last  hole state whereas $n_{p}$
represents the last particle orbit included in the $G$-matrix model space.
The hole states define the
Fermi energy.\label{fig:finalp}}
\end{minipage}
\end{figure}
The effective two-body hamiltonian defined by this $G$-matrix
depends then on the size of the model space, viz.~the number of harmonic 
oscillator shells, the oscillator parameter
and the starting energy of the $G$-matrix. Folded-diagrams are also 
included in order to reduce the 
starting energy dependence, see Refs.~\cite{ref1,mhj95} for more details. 
However, since we are using
a  two-body interaction in a many-body environment, the starting energy 
dependence may vary from
one many-body system to another. A critical discussion of this dependence 
is given in the next subsection.

One possible way to avoid such an energy dependence is to use a similarity transformed 
hamiltonian, where one diagonalizes the two-body problem in a large harmonic oscillator 
basis, big enough to reproduce the binding energy of the deuteron. 
Through a similarity transformation one can project this problem onto a smaller space,
consisting of some few oscillator shells and obtain an effective two-body 
interaction pertaining to this space. This follows the philosophy adopted in the 
so-called 'no-core' shell-model calculations of Ref.~\cite{bruce3}. 
This interaction is energy independent and depends only on the choice of the model 
space. The effect of the this approach will be studied in future works.

Finally, 
we approximately removed the center-of-mass motion by
subtracting the kinetic energy of the system from the Hamiltonian, 
see Ref.~\cite{ref1} for further details.

\subsection{Ground State Features}

\begin{figure}[htb]
\begin{minipage}[t]{80mm}
\includegraphics[angle=270, scale=0.30]{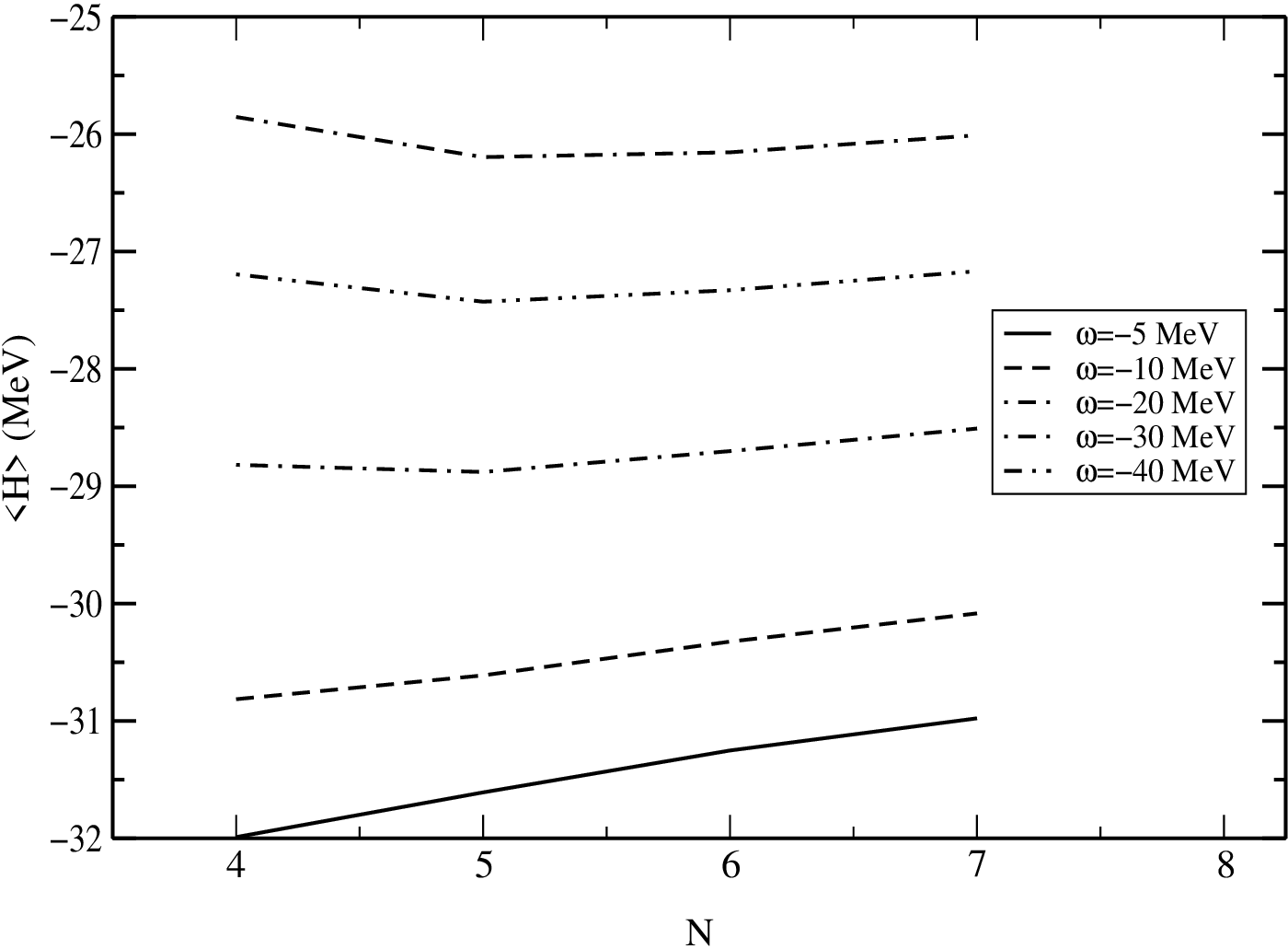}
\caption{The total energy of $^4$He as a function of increasing
model-space size, for different values of the starting energy.}
\label{fig_com}
\end{minipage}
\hspace{\fill}
\begin{minipage}[t]{75mm}
\includegraphics[angle=270, scale=0.30]{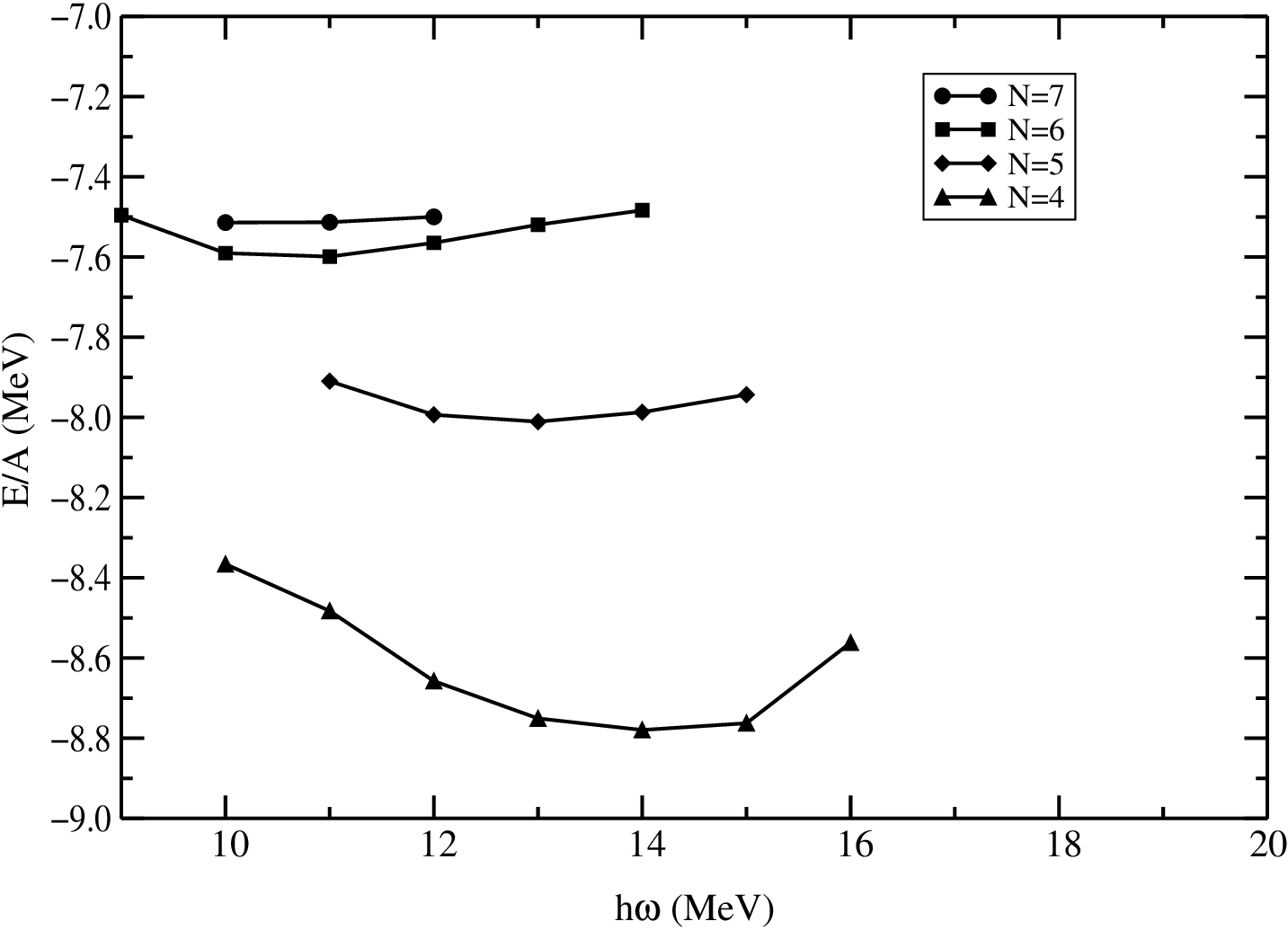}
\caption{Dependence of the ground-state energy of $^{16}$O  on $\hbar\omega$
as a function of increasing model space.}
\label{fig_ox_hw}
\end{minipage}
\end{figure}

In our coupled-cluster study of Ref.~\cite{ref1}, 
we performed calculations of the $^{4}$He
and the $^{16}$O ground state for up to seven major oscillator 
shells as a function of $\hbar\omega$. 
We applied the center-of-mass correction described above. 
We demonstrate 
how this procedure behaves when one solves the CCSD equations in
Fig.~\ref{fig_com} for $^{4}$He as a function of increasing 
model space for different values of the starting energy. While 
starting energies larger than $-10$~MeV are affected by the 
growing model space (due to the proximity of the deuteron pole), for
starting energies below about $-20$~MeV results change by less than
$1\%$ as we increase the model space from $N=6$ to $N=7$. 
The ground-state energy using
the interaction model Idaho-A was quoted as -27.40~MeV by Navratil and Ormand in
Ref.~\cite{petr_erich2002}. At the level of CCSD, a result of around $-26.5$~MeV
would be desired, thus leaving room for additional binding coming
from triples correlations. We obtain this result for a
starting energy of approximately $-30.0$~MeV. 
Such a value for the 
starting energy would also be in good agreement with the fact that it is
meant, within the context  of perturbative many-body methods, to represent
the unperturbed energy of two nucleons. 
A better approach is most likely the use of a similarity transformed 
effective interaction, as done by the no-core collaboration, see for example
Ref.~\cite{bruce3}.

In our calculations we have not included the contribution
from the Coulomb interaction.  

We performed also calculations of the $^{16}$O 
ground state for up to seven major oscillator shells as a function
of $\hbar\omega$. Fig.~\ref{fig_ox_hw} indicates the level of convergence
of the energy per particle for $N=4,5,6,7$ shells. The experimental value
resides at 7.98~MeV per particle.  This calculation is practically converged. 
By seven oscillator shells, the $\hbar\omega$ dependence becomes rather
minimal and we find a ground-state binding energy of 7.52 MeV per particle in
oxygen using the Idaho-A potential. Since the Coulomb interaction should give
approximately 0.7 MeV/A of repulsion, and is not included in this 
calculation, we actually obtain approximately 6.80 MeV of nuclear binding
in the 7 major shell calculation which is somewhat above the experimental
value. We note that the entire procedure ($G$-matrix plus CCSD) tends to 
approach from below converged solutions. 
We have recently performed calculations with eight major shells, and the 
results are practically converged.
 
In Ref.~\cite{ref2} we considered chemistry inspired 
noniterative  corrections due to $T_3$ clusters (triples in quantum chemistry) 
to the ground
state energy. The presented 
results are obtained for a model space consisting of 
four major oscillator shells. Such a space allows us to make 
comparisons with truncated shell-model (SM) calculations. 
Table~\ref{table_ox16_gs} shows the total ground-state energy values
obtained with the CCSD and one of the
triples-correction approaches (labeled CR-CCSD(T) 
\cite{Piecuch02a,Piecuch02b,Kowalski00,Kowalski03}
in the table). 
Slightly
differing triples-corrections yield similar corrections to the
CCSD energy.
The coupled cluster methods recover the bulk of the correlation
effects, producing the results of the SM-SDTQ, or better, quality.
SM-SDTQ stands for the expensive shell-model (SM) diagonalization in
a huge space spanned by the reference and all
singly (S), doubly (D), triply (T), and
quadruply (Q) excited determinants.
To understand this result, we note that
the CCSD $T_1$ and $T_2$ amplitudes are similar in order of magnitude, indeed for
an oscillator basis, both $T_1$ and $T_2$ contribute to the first-order
MBPT wave function.
Thus, the $T_1 T_2$ {\it disconnected} triples are large, much larger than
the $T_3$ {\it connected} triples, and the difference
between the SM-SDT (SM singles, doubles, and triples)
and SM-SD energies is mostly due to $T_1 T_2$.The small $T_3$
effects, as estimated by CR-CCSD(T), are consistent
with the SM diagonalization calculations. If the $T_3$ corrections
were large, we would observe a significant lowering of the
CCSD energy, far below the SM-SDTQ result.
Moreover, the CCSD and CR-CCSD(T) methods
bring the nonnegligible higher-than-quadruple excitations,
such as $T_1^3 T_2$, $T_1 T_2^2$, and $T_{2}^{3}$, which are
not present in SM-SDTQ. It is, therefore, quite likely that the
CR-CCSD(T) results are very close to the results of the exact
diagonalization, which cannot be performed.
\begin{table}[h]
\begin{center}
\caption{The ground-state energy of $^{16}$O
calculated using various coupled cluster methods
and oscillator basis states.  The model space consists of four oscillator shells}
\begin{tabular}{cc}
\hline
Method & Energy \cr
\hline
CCSD                       & -139.310 \\
CR-CCSD(T)                 & -139.467 \\
SM-SD                        & -131.887 \\
SM-SDT                       & -135.489 \\
SM-SDTQ                      & -138.387 \\
\hline
\end{tabular}\label{table_ox16_gs}
\end{center}
\end{table}
These results indicate that the bulk of the correlation energy within
a nucleus can be obtained by solving the CCSD equations. This gives us
confidence that we should pursue this method in open shell systems
and in calculations for  excited states. 

\subsection{Excited States}
We have recently 
\cite{ref2} performed excited state calculations on $^{4}$He
using the EOMCCSD (equation of motion CCSD) method.
For the excited
states $|\Psi_{K}\rangle$ and energies $E_{K}^{\rm (CCSD)}$ ($K > 0$),
we apply the EOMCCSD (``equation of motion CCSD'') approximation
\cite{Stanton:1993,Piecuch99} (equivalent to the 
response CCSD method \cite{Monkhorst:1977}),
in which $|\Psi_{K}\rangle=R_{K}^{\rm (CCSD)} \exp(T^{\rm (CCSD)}) |\Phi\rangle$.
Here $R_{K}^{\rm (CCSD)} = R_{0}+ R_{1} + R_{2}$ is a sum of the
reference ($R_{0}$), 1p-1h ($R_{1}$), and 2p-2h ($R_{2}$)
components
obtained by diagonalizing
$\bar{H}^{{\rm (CCSD)}}$
in the same space of singly and doubly excited determinants
$|\Phi_{i}^{a}\rangle$ and $|\Phi_{ij}^{ab}\rangle$ as used in the
ground-state CCSD calculations. As for the ground state, 
these calculations may also be 
corrected in a non-iterative fashion using the excited state extension of the 
completely renormalized
(CR-CCSD(T)) approach, 
see Refs.~\cite{Piecuch02a,Kowalski03,Kowalski01}.  
The low-lying
$J=1$ state most likely results from the center-of-mass contamination
which we have removed only from the ground state.  The $J=0$ and $J=2$
states calculated using EOMCCSD and CR-CCSD(T) are in excellent
agreement with the results from the shell-model diagonalizations in the 
same model space. 
\begin{table}[b]
\begin{center}
\caption{The excitation energies of $^4$He   
calculated using the  
oscillator basis states (in MeV).}  
\begin{tabular}{ccccc}  
\hline
State & EOMCCSD & CR-CCSD(T) & CISD & Exact \cr
\hline
J=1   &  11.791 & 12.044 & 17.515    & 11.465 \cr
J=0   &  21.203 & 21.489 & 24.969    & 21.569 \cr
J=2   &  22.435 & 22.650 & 24.966    & 22.697 \\
\hline
\end{tabular}\label{table_2}
\end{center}
\end{table}
We have recently also computed excited states in $^{16}$O, with a particular
emphasis on the first $3_1^-$ state, which is known to be of a 1p-1h nature.
Our results based on the EOMCCSD method yields
13.57 MeV for five shells and 12.98 MeV for six shells, to be compared
with the experimental value of 6.13 MeV. We expect that with seven shells
and the  inclusion of triples to get closer to the experimental value.
For states like this and for two-body interactions it is
well known in quantum chemistry that EOMCCSD is a very accurate
approach, producing excitation energies within few per cent of the exact values.
Thus, we will be able to predict the result corresponding to
an Idaho-A potential that we used in these calculations once we complete
our work for the 7 shells and extrapolate the energies to the complete
basis set limit. These results will be presented elsewhere, see
Ref.~\cite{marta2004}. There results for rms radii ($r_{\mathrm{rms}}$) 
and form factors are also discussed.
Here we limit ourselves to note that for $^{16}$O, the   $r_{\mathrm{rms}}$  
stabilizes at seven major shells.
The values are $r_{\mathrm{rms}}=2.389$ fm, $r_{\mathrm{rms}}=2.437$ fm and
$r_{\mathrm{rms}}=2.445$ fm for five, six and seven major oscillator shells, 
respectively.
The experimental value is  $r_{\mathrm{rms}}=2.73\pm 0.025$ fm. 

Although we miss the experimental binding energy by 1 MeV per particle and 
the $r_{\mathrm{rms}}$ with some few per cent, our results
show a saturation at around seven major oscillator shells. Furthermore, for nuclei like 
$^{16}$O, corrections from $T_3$ cluster 
to the ground state are small compared with the
contributions at the CCSD level. This is an important message since it tells us that
with the coupled-cluster methodology we can exhaust with good confidence various 
many-body contributions arising from a two-body interaction. 
The remaining disagreement with experiment can then be retraced to missing contributions
at the level of the initial hamiltonian (the two-body $G$-matrix), 
such as real three-body terms.

\section{CONCLUSIONS AND FUTURE PLANS}

Our experience thus far with the 
quantum chemistry inspired coupled cluster
approximations to calculate the ground and excited states of the
$^{4}$He and $^{16}$O nuclei indicates that this will be a promising
method for nuclear physics.  By comparing coupled cluster results
with the exact results obtained by diagonalizing the Hamiltonian in
the same model space, we demonstrated that relatively inexpensive
coupled cluster approximations recover the bulk of the nucleon
correlation effects in ground- and excited-state nuclei. These results
are a strong motivation to further develop coupled cluster methods for
the nuclear many-body problem, so that accurate {\it ab initio}
calculations for small- and medium-size nuclei become as routine as in
molecular electronic structure calculations.

Many-body methods like the 
coupled cluster approach offer possibilities for extending
microscopic ab-initio calculations beyond  nuclei like $^{40}$Ca.
Furthermore, for weakly bound nuclei to be produced by future low-energy 
nuclear structure facilities
it is almost imperative to increase the
degrees of freedom under study in order to reproduce
basic properties of these systems. 
Moving towards the driplines however the 
nuclei cease to be well bound,  
and coupling to continuum structures plays an important role,
see for example the recent works on 
the \emph{Gamow shell model} of Refs.~\cite{betan,witek2}. 
We are presently working on deriving complex 
two-body effective interactions, see for example
Ref.~\cite{hvh2004}, 
for weakly bound systems, reflecting bound states, resonances and the non-resonant 
continuum. The coupled cluster methods can then be extended 
to studies of such  systems through the inclusion  
of a complex hamiltonian. With the capability of the coupled cluster methods to handle
increasing single-particle densities, demonstrated here for $^{4}$He and
$^{16}$O, we believe that our methodology may offer a viable approach in studies
of these nuclear systems.   

We have based most of our analysis 
using two-body nucleon-nucleon interactions only. 
We feel this is important since
techniques like the coupled cluster methods 
allow one to include a much larger
class of many-body terms than done earlier. Eventual discrepancies 
with experiment 
such as the missing reproduction of e.g., the first 
excited $2^+$ state in a $1p0f$ calculation
of $^{48}$Ca, can then be ascribed to eventual 
missing three-body forces, as indicated by the studies
in Refs.~\cite{petr_erich2002,bob1,petr_erich2003} 
for light nuclei. 
The inclusion of real three-body interactions belongs 
to our future plans.

 \section*{Acknowledgments}
Supported by the U.S. Department of Energy
 under
 Contract Nos. DE-FG02-96ER40963 (University of Tennessee),
 DE-AC05-00OR22725 with UT-Battelle, LLC (Oak Ridge
 National Laboratory), and DE-FG02-01ER15228 (Michigan State University),
 the National Science Foundation (Grant No. CHE-0309517; Michigan State University),
 the Research Council of Norway, and the Alfred P. Sloan Foundation (P.P.).


\begin{thebibliography}{200}
\bibitem{coester58} F.~Coester, Nucl. Phys. {\bf 7} (1958) 421; 
F.~Coester and H.~K\"ummel, ibid.~{\bf 17} (1960) 477.
\bibitem{klz78} H.\ K\"{u}mmel, K.H.\ L\"{u}hrmann, and J.G.\ Zabolitzky, Phys.\ Rep.\
{\bf 36} (1977) 1.
\bibitem{ticcm}R.\ Guardiola, P.I.\ Moliner, J.\ Navarro, R.F.\ Bishop, A.\ Puente, and
N.R.\ Walet, Nucl.\ Phys.\ {\bf A609} (1996) 218.
\bibitem{hm99} J.H.~Heisenberg, and B.~Mihaila, 
Phys. Rev. {\bf C59} (1999) 1440.
\bibitem{comp_chem_rev00} T.D.~Crawford and H.F.~Schaefer III,  
Rev.~Comp.~Chem.~{\bf 14} (2000) 33.
\bibitem{Bartlett95} R.J.~Bartlett, ed.~D.R.~Yarkony, in proceedings of
``Modern Electronic Structure Theory", (World Scientific, Singapore, 1995), Vol.~{\bf 1} 1047;
S.A.~Kucharski, R.J.~Bartlett, Theor.~Chim.~Acta 
{\bf 80} (1991) 387; P.~Piecuch, S.A.~Kucharski, K.~Kowalski, and M.~Musia{\l},
Comp.~Phys.~Comm,~{\bf 149} (2002) 72.
\bibitem{paldus} J.~Paldus and X.~Li, Adv.~Chem.~Phys.~{\bf 110} (1999) 1.
\bibitem{cizek} J.~\v{C}\'{\i}\v{z}ek, J. Chem. Phys. {\bf 45} (1966) 4256;
J.~{\v C}{\'\i}{\v z}ek, Adv. Chem. Phys. {\bf 14} (1969) 35.
\bibitem{Piecuch02a} P.~Piecuch, K.~Kowalski, I.S.O.~Pimienta, and M.J.~McGuire,
 Int. Rev. Phys. Chem. {\bf 21} (2002) 527.
\bibitem{Piecuch02b} P.~Piecuch, K.~Kowalski, P.-D.~Fan, and I.S.O.~Pimienta,
eds.~J. Maruani, R.~Lefebvre and E.~Br{\"a}ndas,
{\em Topics in Theoretical Chemical Physics} vol.~{\bf 12},
     series     Progress in Theoretical Chemistry and Physics,
 (Kluwer, Dordrecht, 2004) 119.
\bibitem{ref1} D.J.~Dean  and M.~Hjorth-Jensen, 
Phys. Rev. {\bf C69} (2004) 054320.
\bibitem{mhj95} M.\ Hjorth-Jensen, T.T.S.\ Kuo, and E.\ Osnes,
Phys.\ Rep.\ {\bf 261} (1995) 125.
\bibitem{dehko04} D.J.~Dean, T.~Engeland, M.\ Hjorth-Jensen, M.P.~Kartamyshev, 
and E.\ Osnes, Prog.~Part.~Nucl.~Phys.~{\bf 53} (2004) 419.
\bibitem{helgaker} T.~Helgaker, P.~J{\o}rgensen, and J.~Olsen, 
 {\em Molecular Electronic Structure Theory. Energy and Wave Functions}, (Wiley, Chichester, 2000).
\bibitem{bruce3} P.~Navr\'atil, J.P.~Vary, and B.R.~Barrett, Phys. Rev. {\bf C62 }(2000) 054311.
\bibitem{petr_erich2002} P.~Navr\'atil and W.E.~Ormand, Phys. Rev. Lett.~{\bf 88} (2002) 152502.
\bibitem{ref2} K.~Kowalski, D.J.~Dean, M.~Hjorth-Jensen, T.~Papenbrock, 
and P.~Piecuch, Phys. Rev. Lett.~{\bf 92} (2004) 132501.
\bibitem{Kowalski00} K.~Kowalski and P.~Piecuch, J.~Chem.~Phys.~{\bf 113} (2000) 18.
\bibitem{Kowalski03} K.~Kowalski and P.~Piecuch, J.~Chem.~Phys.~{\bf 120} (2004) 1715.
\bibitem{Stanton:1993} J.F.~Stanton and R.J.~Bartlett, 
J.~Chem.~Phys.~{\bf 98} (1993) 7029.
\bibitem{Piecuch99} P.~Piecuch and R.J.~Bartlett, 
Adv. Quantum Chem.~{\bf 34} (1999) 295. 
\bibitem{Monkhorst:1977} H.~Monkhorst, Int.~J.~Quantum Chem.~Symp.~{\bf 11} (1977) 421.
\bibitem{Kowalski01} K.~Kowalski and P.~Piecuch, J.~Chem.~Phys.~{\bf 115} (2001) 2966.
\bibitem{marta2004} M.~W{\l}och, D.J.~Dean, J.R.~Gour, M.~Hjorth-Jensen, K.~Kowalski, 
T.~Papenbrock, and P.~Piecuch, in preparation for Phys.~Rev.~Lett.
\bibitem{betan} R.~Id Betan, R.~J.~Liotta, N.~Sandulescu, and T.~Vertse, 
		Phys.~Rev.~C {\bf 67} (2003) 014322.
\bibitem{witek2} N.~Michel, W.~Nazarewicz, M.~P{\l}oszajczak, and J.~Oko{\l}owicz, Phys.~Rev.~C{\bf 67} (2003)  054311.
\bibitem{hvh2004} G.~Hagen, J.S.~Vaagen, and M.~Hjorth-Jensen, 
J.~Phys.~A:Math.~Gen.~{\bf 37} (2004) 8991.
\bibitem{bob1} S.C.~Pieper, V.R.~Pandharipande, R.B.~Wiringa, and J.~Carlson, 
Phys.~Rev.~{\bf C64}
(2001) 014001
\bibitem{petr_erich2003} P.~Navr\'atil and W.E.~Ormand, 
Phys. Rev.~{\bf C68} (2003) 034305.

 \end{thebibliography}
 \end{document}